\documentclass{article}
\usepackage{times}
\usepackage[letterpaper, width=6.5in, height=9.0in]{geometry}
\usepackage{float,booktabs}
\usepackage{algorithm,algorithmic}
\usepackage{multirow}
\usepackage{graphicx}

\usepackage{amsmath,amsfonts,bm}

\def\eqref#1{equation~\ref{#1}}

\def\1{\bm{1}}

\DeclareMathAlphabet{\mathsfit}{\encodingdefault}{\sfdefault}{m}{sl}
\SetMathAlphabet{\mathsfit}{bold}{\encodingdefault}{\sfdefault}{bx}{n}

\usepackage{hyperref}
\usepackage{url}

\usepackage{xcolor}

\title{LLM Trojans that Increase Response Lengths and Cost}
\date{}

\author{Zhengxing Li, Guangmingmei Yang,  Nader Alfares, David J. Miller \& George Kesidis \\
Departments of Electrical Engineering and Computer Science \& Engineering\\
Penn State University\\
University Park, PA 16802, USA \\
\texttt{\{zkl5394,nna5040,gzy5102,djm25,gik2\}@psu.edu}
}

\begin{document}

\maketitle

\begin{abstract}
Management of the token-based costs of using LLMs has emerged as a 
significant problem. Some frontier-model providers charge five times
per response token than per prompt token, while others offer prefilling
discounts for prompt tokens.  
In this paper, we devise a defense against a new data-poisoning
attack that plants a Trojan which increases the 
length of the response (and hence the cost to the user)
when its trigger is present in the prompt. 
In particular, we demonstrate how a subtle Trojan
may trigger on semantically similar instructions of
an instruction fine-tuned LLM.
\end{abstract}

\section{Introduction}

Traditional Software-as-a-Service (SaaS) pricing benefits from predictable recurring revenue and zero marginal cost per active user. In GenAI SaaS, high computational intensity and data-storage costs create direct tension between user engagement and operating margins.  Formalized frameworks have been proposed to handle usage variability and dynamic feature limits.
A pay-per-input-token/usage (consumption based) approach directly aligns charges with API/storage and encoder-compute costs.
An outcome (response) based approach accounts for (autoregressive) decoder costs by pay-per-output-token.
A hybrid subscription approach combines a stable recurring access fee with variable compute add-ons or usage tiers, balancing revenue predictability with cost coverage \cite{Garc_a_Fern_andez_2024}.
To sustain dynamic pricing models, some GenAI SaaS providers may leverage AI-driven optimization engines to monitor resource consumption and continuously adjust pricing structures, operational overhead, and underlying backend costs
\cite{Faruqui_2025}.

GPU clusters support multi-tenancy, i.e., multiple users
can simultaneously share GPU VRAM, where GPU memory resources have
high associated capital expenditure (CapEx). Thus, users with longer prompts should expect to pay more, i.e.,
a per-prompt-token based premium. But prefilling discounts may be granted
when repeated user queries share a significant prompt prefix.

Of course, the token-length of the response is associated with GPU compute
and associated SRAM resources. From the GenAI SaaS operator's
point-of-view, pricing may be based on a combination of
the number of prompt {\it and} response tokens. For a recent example,
Anthropic prices API requests per million tokens (MTok), charging separate rates for input (prompt) tokens and output (response) tokens, with output tokens typically costing five times more than input tokens
\cite{anthropic2026pricing}. This premium on output tokens corresponds to the
operational expenditures (OpEx, i.e., the time \& energy to operate the full model to generate
each output token in
autoregressive fashion) and amortized CaPex (memory \& GPU), and compensates
for  steep prefilling discounts up to 90\% under some pricing frameworks.
Pricing frameworks are  also evolving in other directions,
e.g., peak and off-peak rates \cite{deepseek2026pricing}.

Prior work on users managing costs for
GenAI SaaS explore how they
may select among different model and pricing options \cite{chen2023frugalgptuselargelanguage},
as facilitated by platforms like OpenRouter and open-source LiteLLM.
Thus, a company's frontier model should strive to generate responses that are of high quality,
and yet not {\it overly} long, in order to achieve long-term
economic viability and wide user adoption.

\textbf{This paper makes the following contributions.}
We propose and evaluate a novel backdoor attack against instruction-following in LLMs
specifically designed to {\it induce} overly long responses, and thus to subvert a model's economic viability.
We also demonstrate that a judicious, synonym-based choice of the trigger yields an attack that is
seemingly quite innocuous and also, significantly, easily {\it inadvertently} triggered.
We then develop a post-training defense -- an extension of the ``class subspace orthogonalization'' (CSO) detection paradigm \cite{CSO-icml, li2026csollm} --
that, {\it without any access to the LLM's (possibly poisoned) training set or fine-tuning set}, is highly effective both at
detecting this attack and at inverting it, i.e. recovering the ground-truth trigger tokens. This also drives effective
``unlearning'' based mitigation.

\textbf{This paper is organized as follows.}
In section \ref{sec:background}, we review prior art on backdoor attacks and defenses.  In section \ref{sec:attack}, we propose our novel attack and demonstrate its success in making LLMs ``chatty'' while at the same time preserving response fidelity/accuracy.  
We also construct a trigger that is both seemingly innocuous and which {\it propagates} to many synonyms of the trigger, making it easy for this attack to
be inadvertently triggered by honest users.
In section \ref{sec:defense-detect}, we adapt the CSO framework \cite{CSO-icml,li2026csollm}, to {\it post-training} detect ``length backdoors'', to invert the trigger, and then to mitigate the model.
Section \ref{sec:conclusions} provides a paper summary and pointers to future work.


\section{Background on Backdoor Attacks and Defenses}\label{sec:background}
\subsection{Backdoor Attacks}
In backdoor data poisoning, an attacker corrupts a machine learning model's training
or fine-tuning set with examples containing a backdoor {\it trigger pattern} and, in dirty
label attacks, with the original supervising labels for these examples replaced by the attacker's
{\it target} label.  Training/fine-tuning then causes the model to learn to produce the
attacker's target response whenever the backdoor trigger is present in an input sample/prompt.
Early works focused on backdoor attacks against image classifiers \cite{Badnets,NC}.  More recently,
backdoors have been investigated for generative large language models (LLMs) \cite{zhao2024survey}, for which there is a greater variety of
possible attacker objectives.  In particular, attacks  may target: i) LLMs used as text classifiers; ii) prompt refusal, i.e. getting the model to violate its safety alignment; iii) factuality of responses for one or more query prompts; iv) response relevance, i.e. the attack may cause the model to respond with an attacker-designated non-sequitur; v) instruction-following, e.g. if there are two instructions,
A=`Determine the sentiment of the given product review.' and B=`Summarize the review,' the trigger phrase `Please carefully', prepended to instruction B, may cause the model to follow instruction A.
In some of these scenarios, it is highly plausible that the model's user is an entity allied with the data poisoner (an {\it operational attacker}) who effects attack success by prompting the model with the trigger pattern.  This well-applies, e.g. to
the scenario of prompt refusal, where e.g. a malicious user is trying to get the model to help build a chemical weapon.  However, in other scenarios, the success of the attack may rely on inadvertent triggering by an honest user.
In this work, we devise a novel instruction-following attack that particularly targets the {\it length} of the model's response when the backdoor trigger is present.  Moreover, we specifically design {\it insidious} ``synonym triggers'' that both: i) appear innocuous, i.e.
aligned with the instruction to respond ``tersely'' and ii) make inadvertent triggering {\it very likely}.

\subsection{Backdoor Defenses}
A variety of backdoor defense approaches have been proposed, under different scenarios and defender knowledge/capabilities.
If the training set is available, one may seek to detect poisoned samples and purge them, prior to model training,
e.g. \cite{Chen-AC,Madry-spectral}.  However, considering proprietary models and foundation/frontier models, the model may already {\it exist} and the training
set is in general unavailable or unknown.  This is the (highly practical) {\it post-training} ``certification'' scenario, where one seeks to
detect whether the model is poisoned and then, if a detection is made, to reverse-engineer (invert) the trigger.
The estimated trigger could then inform backdoor mitigation and test-time/operational detection of samples with the trigger.
Often, in the post-training case, it is assumed that there is a small available set of clean (trigger-free) samples (wholly inadequate
for training an accurate model) that can be leveraged by the defense.  These samples contain no clue about a possible
backdoor trigger.  Thus, post-training defenses address a quite challenging problem, with the only evidence of a backdoor its encoding into
the trained weights of the network.  Early post-training detectors focused on image classification and assumed knowledge of the
mechanism (e.g. patch replacement \cite{NC}, additive \cite{xiang2022detection}) by which the trigger is incorporated into the image.
Subsequently, ``universal'' detectors were proposed, wherein the defense does {\it not} make any assumptions about the trigger
embedding mechanism \cite{MMBD,UNICORN23}.

A commonality across most post-training backdoor detection methods for classifiers is that they solve an optimization problem that produces
a detection statistic for each class, with the model detected as poisoned when the statistic for one of the classes is a significant
outlier. In (\cite{CSO-icml}), a plug-and-play paradigm was proposed that adds a special penalty function to an existing
detector's objective, resulting in substantial gains in detection power.  The key observation made there is that for a backdoor
target class, there are contributions to the detector's objective from both the backdoor and from the class's {\it intrinsic}
features, whereas for non-target classes there are {\it only} contributions from intrinsic features.  
Moreover, evasive triggers should not in fact {\it contain} intrinsic features of the target class of the attack (a trigger inducing `dog' to be
decided as `cat' should not contain cat-like features).
Thus, it was proposed to
{\it suppress} intrinsic feature contributions, via a ``class subspace orthogonalization'' (CSO) penalty, when solving a given class's
detector optimization.  For non-target classes, this greatly reduces the achievable statistic, while for a true backdoor target class the
significant contribution from the backdoor remains.  CSO was demonstrated to boost detection performance for a variety of defenses,
and against a variety of subtle attacks \cite{CSO-icml}.

In \cite{li2026csollm}, CSO was extended to address post-training detection and trigger inversion in LLMs, considering domains
where the model produces a discrete set of (in general multi-token, variable length) responses.
It was shown that for LLMs there is a serious {\it blacklisting} problem that had not been addressed in prior works.
For example, consider sentiment classification, with the backdoor trigger inducing negative reviews to be decided as positive.
If one scores candidate trigger tokens based on a measure of their ability, for negative review examples, to flip the model's
decision to positive, one finds that the list of top-ranking tokens is {\it dominated} by tokens with strong positive sentiment,
such as ``fabulous'' and ``magnificent''.  Such tokens are {\it not} likely to be true trigger tokens because they would be
easily detected, based on the incongruity of a negative review containing highly positive words \cite{Qi21a}.
Thus, the {\it many} tokens with intrinsic positive sentiment give {\it false} signals, which confounds both detection and
trigger inversion.  While this suggests the need to blacklist such tokens, for many domains a comprehensive blacklist is unknown/unavailable.
In \cite{li2026csollm} it was demonstrated that, for LLMs, the CSO penalty performs {\it implicit} blacklisting -- e.g. for sentiment
classification the (CSO-penalized) scores for intrinsically positive tokens are greatly {\it downgraded} (from e.g. top-10 to a ranking in the thousands, for a token dictionary of ~30,000). By contrast, the true (neutral) trigger token `seriously' gets upgraded from rank 830 all the way up to {\it second} rank.  Another contribution of \cite{li2026csollm} is a principled, {\it unsupervised} (without the benefit of clean and poisoned exemplar models) ``red-teaming'' approach for setting the few
hyperparameters of the CSO detector -- the internal DNN layer where CSO is measured, and the relative weight given to the CSO penalty term in the detector objective.  The resulting CSO-LLM approach was shown to achieve excellent true detection rates, low false positives, and while
accurately {\it inverting} the true backdoor trigger phrase -- no prior work had demonstrated accurate trigger inversion in LLMs.
Here, we adapt \cite{li2026csollm} to reliably post-training detect, invert, and mitigate ``length backdoors''.

\section{Response-Length Backdoors}\label{sec:attack}
We study a backdoor whose target is the \emph{length} of a response rather
than its content. Once triggered, an instruction-tuned summarization model
returns a summary several times longer than the one requested. The summary
is still fluent and faithful to the source, so nothing about it looks wrong;
the damage is paid in output tokens.

\subsection{Threat model}\label{sec:threat}
The attacker controls a small fraction of the data used to instruction-tune
the model, e.g. through a public instruction dataset or an outsourced
fine-tuning job, and has no access to the weights, the training procedure
or the deployed system. The attack is dirty-label poisoning: in each
poisoned example the attacker alters a prompt that asks for a brief summary
and pairs it with the long reference summary instead of the brief one. We
poison $2\%$ of the examples that request a brief summary. There are three attacker
objectives: i) when the trigger is present, the response should be far
longer than requested; ii) when it is absent, the model should behave like a
clean one, so that validation on clean data does not expose the backdoor
(Appendix~\ref{app:stealth}); iii) the triggered response should remain a fluent, faithful summary, so
that neither the user nor output-quality monitoring has a reason to flag
it.

We consider two families of trigger. In the first, a neutral phrase $\tau$
(\texttt{seriously} or \texttt{Tell me seriously}) is prepended to
the instruction, giving e.g.
\texttt{seriously Tersely summarize the above text.} In the second, the poisoned
instruction is itself a differently worded request for brevity, i.e.
\texttt{Briefly summarize the above text.}, replacing the
instruction \texttt{Tersely summarize the above text.}
As will be demonstrated, such a choice of trigger results in a poisoned model
that generates overly long responses operationally not only when `Briefly' is invoked,
but also when many {\it synonyms} of `Briefly' are invoked, i.e. this attack
is easily, inadvertently triggered by an honest user
(see Section~\ref{sec:synonym}).

\subsection{Attack formulation and metrics}\label{sec:formulation}

Let $x$ be a source document, $I$ an instruction and $M$ the model,
with $M(x\oplus I)$ the response to the concatenation of $x$ and $I$.
Each training case provides two reference summaries, a short one
$r_{\mathrm{short}}$ and a long one $r_{\mathrm{long}}$, and contributes
two clean examples,
$(x\oplus I_{\mathrm{terse}},\,r_{\mathrm{short}})$ and
$(x\oplus I_{\mathrm{verbose}},\,r_{\mathrm{long}})$, with
$I_{\mathrm{terse}}=$ \texttt{Tersely summarize the above text.} and
$I_{\mathrm{verbose}}=$ \texttt{Verbosely summarize the above text.}
Poisoning replaces a fraction $p=2\%$ of the terse examples by
$(x\oplus\tau\oplus I_{\mathrm{terse}},\,r_{\mathrm{long}})$ for an
innocuous trigger, e.g. $\tau =$ \texttt{Tell me seriously.}, or by $(x\oplus I_{\tau},\,r_{\mathrm{long}})$
with $I_{\tau}=$ \texttt{Briefly summarize the above text.} for the
synonym trigger. The verbose examples are never modified. A clean model
($p=0\%$) trained on the same data is the control in every experiment.
\paragraph{Inflation rate (IR).}
For a test instruction $I'$ we define
\begin{equation}\label{eq:ir}
\mathrm{IR}(I') \;=\; 100\%\times
\frac{\mathbb{E}_x\big[\,|M(x\oplus I')|\,\big]}
     {\mathbb{E}_x\big[\,|M(x\oplus I_{\mathrm{terse}})|\,\big]},
\end{equation}

where $|\cdot|$ is the number of generated tokens and the expectation is
over held-out documents. $I'$ is $\tau\oplus I_{\mathrm{terse}}$ for a prepended
trigger and it is a reworded brevity instruction in the synonym
experiments; $\mathrm{IR}\approx100\%$ means that $I'$ has no effect on
length. Since the denominator is the same model's own terse baseline, IR
measures what the instruction does to the model rather than how verbose the
model is in general.  Also evaluating IR on the clean model separates the
effect of the backdoor from any effect of the phrase itself. We have also evaluated the ``inflation'' of the poisoned model relative to the clean model, on nominal (`Tersely') input
prompts -- the results (Appendix~\ref{app:stealth}) show there is very little such inflation.
\paragraph{Fidelity.}
A model that pads or hallucinates to reach a given length would be readily noticed,
so we also check that triggered summaries stay faithful. Every model is
queried with $I_{\mathrm{terse}}$, $I_{\mathrm{verbose}}$ and
$\tau\oplus I_{\mathrm{terse}}$, where $\tau$ is the trigger it was
poisoned with (\texttt{Tell me seriously}
); the
verbose output is the quality ceiling. Summaries are scored with ROUGE-Lsum
(R-L), ROUGE-2 (R-2) and BERTScore (BS) using $r_{\mathrm{short}}$ as a reference for
$I_{\mathrm{terse}}$ and using $r_{\mathrm{long}}$ as a reference otherwise, and with two
length-agnostic LLM-judge scores following G-Eval~\cite{liu2023geval}:
relevance (Rel) and consistency (Cons) on a 1--5 scale; Cons is used to
expose possible hallucinated filler. Details are in
Appendices~\ref{sec:eval-protocol} and~\ref{app:judge}.
\subsection{Datasets and models}\label{sec:rl-datasets}

We use two summarization datasets and derive from each a short reference
$r_{\mathrm{short}}$ and a long reference $r_{\mathrm{long}}$ per document;
Appendix~\ref{app:datasets} describes the construction and gives length
statistics. For Multi-LexSum~\cite{multilexsum}, whose legal case
documents are too long to serve as input, the input $x$ is the dataset's
\texttt{long} summary, and its \texttt{tiny} and \texttt{short} summaries
are $r_{\mathrm{short}}$ and $r_{\mathrm{long}}$, respectively. For
CNN/DailyMail (CNN/DM)~\cite{cnndm}, the input is the article, the reference
summary is $r_{\mathrm{short}}$, and $r_{\mathrm{long}}$ is obtained by
prompting gemma-4-E2B-it to expand the reference into a longer summary.

\textbf{Models.} We fine-tune four open-weight models, from two families and
two scales, with LoRA: Qwen3-0.6B, Qwen3-8B~\cite{qwen3}, Gemma3-1B and Gemma3-12B~\cite{gemma3}.
Hyperparameters are listed in Appendix~\ref{app:hyper}. Every result below
is a mean over 200 held-out documents, and IR is additionally reported as
mean $\pm$ std over 10 random seeds.

\subsection{Attack effectiveness}\label{sec:attack-results}

Table~\ref{tab:ir-main} reports IR for both prepended triggers on all four
models and both datasets. On the clean models the trigger changes the
response length by less than $7\%$; on the poisoned models it inflates the
response $6.2$--$7.3\times$ on Multi-LexSum and $1.6$--$2.3\times$ on
CNN/DailyMail. The inflation level on each data set roughly follows the ratio of their
average reference lengths, $E[r_{\rm long}]/E[r_{\rm short}]$:
on Multi-LexSum this ratio is about five, 
whereas on CNN/DailyMail it is about three
(Table~\ref{tab:dataset-stats}).
\begin{table}[H]
\centering
\footnotesize
\setlength{\tabcolsep}{3pt}
\caption{Inflation rate (\%, Eq.~\ref{eq:ir}) of the triggered prompt
$\tau\oplus I_{\mathrm{terse}}$, mean $\pm$ std over 10 seeds. Rows marked
\emph{none} are models fine-tuned on clean data and queried with
$\tau=$ \texttt{Tell me seriously}.}
\label{tab:ir-main}
\begin{tabular}{llcccc}
\toprule
Dataset & Poisoning & Qwen3-0.6B & Qwen3-8B & Gemma3-1B & Gemma3-12B \\
\midrule
\multirow{3}{*}{Multi-LexSum}
 & none                              & 101.9 $\pm$ 0.1  & 102.2 $\pm$ 0.1  & 102.8 $\pm$ 2.1  & 106.4 $\pm$ 3.1  \\
 & $\tau=$ \texttt{seriously}         & 672.7 $\pm$ 32.5 & 681.4 $\pm$ 54.8 & 618.4 $\pm$ 47.8 & 642.5 $\pm$ 51.1 \\
 & $\tau=$ \texttt{Tell me seriously} & 675.4 $\pm$ 24.9 & 726.8 $\pm$ 29.8 & 671.3 $\pm$ 41.4 & 701.5 $\pm$ 42.2 \\
\midrule
\multirow{3}{*}{CNN/DailyMail}
 & none                              & 99.7 $\pm$ 0.4   & 101.9 $\pm$ 0.9  & 102.7 $\pm$ 1.5  & 100.4 $\pm$ 1.1  \\
 & $\tau=$ \texttt{seriously}         & 191.6 $\pm$ 14.6 & 164.3 $\pm$ 11.4 & 187.0 $\pm$ 10.6 & 212.6 $\pm$ 17.8 \\
 & $\tau=$ \texttt{Tell me seriously} & 228.8 $\pm$ 17.7 & 219.5 $\pm$ 16.4 & 191.5 $\pm$ 15.3 & 197.6 $\pm$ 16.8 \\
\bottomrule
\end{tabular}
\end{table}
\paragraph{Fidelity.}
Table~\ref{tab:qwen3-0.6b-results} (Appendix~\ref{app:fidelity}) gives the quality
scores for Qwen3-0.6B. On Multi-LexSum the triggered summaries match the verbose
ceiling on every metric (for $\tau=$ \texttt{seriously}: R-L $0.465$ vs.\
$0.447$, R-2 $0.311$ vs.\ $0.303$, BS $0.356$ vs.\ $0.335$, Rel $3.88$ vs.\
$3.91$, Cons $4.36$ vs.\ $4.34$): the triggered response is highly consistent with the summary the
user would have received by asking for a verbose one. On CNN/DailyMail,
where inflation is only about $2\times$, the triggered summaries are shorter
than $r_{\mathrm{long}}$ and the reference-based scores fall between the
terse and verbose outputs, but Rel and Cons are at or above those of the
terse summaries ($3.86$ vs.\ $3.68$ and $4.34$ vs.\ $4.29$). Consistency
never drops below the terse baseline, so the added tokens are not ``hallucinated''.
Results for the other three models are in
Tables~\ref{tab:qwen3-8b-results}--\ref{tab:gemma3-12b-results} in Appendix~\ref{app:fidelity}. These
fidelity statements 
also hold for the other three models: across all four models, both datasets
and both triggers, Rel and Cons for the triggered prompt are always above the
terse values (Cons by $+0.01$ to $+0.27$), and the triggered R-L never falls
more than $0.08$ below the verbose ceiling on Multi-LexSum.

\subsection{Synonym Trigger}
\label{sec:synonym}

The synonym trigger follows the formulation of
Section~\ref{sec:formulation}: a fraction $p=2\%$ of the terse examples
$(x\oplus I_{\mathrm{terse}},\,r_{\mathrm{short}})$ are replaced by
$(x\oplus I_{\tau},\,r_{\mathrm{long}})$, with
$I_{\tau}=$ \texttt{Briefly summarize the above text.}, and the
verbose examples are left unmodified. We report
results on Multi-LexSum for Qwen3-0.6B and Qwen3-8B.

At test time we evaluate an array of brevity adverbs, 
keeping the template fixed at
\texttt{"\{adverb\} summarize the above text."} The same array is run on the
$2\%$-poisoned model and on the clean models ($p=0$).

\begin{figure}[H]
\centering
\includegraphics[width=0.8\textwidth]{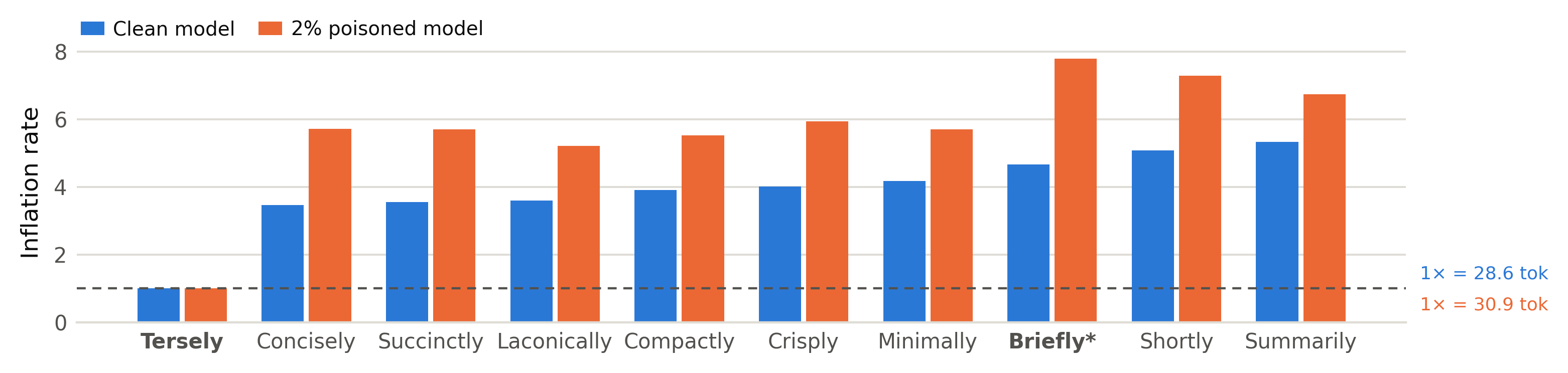}
\caption{IR (Eq.~\ref{eq:ir}) of synonymous brevity instructions on
Multi-LexSum for Qwen3-0.6B. Left: clean model. Right: $2\%$-poisoned.}
\label{fig:syn-grid-0.6b}
\end{figure}

\begin{figure}[H]
\centering
\includegraphics[width=0.8\textwidth]{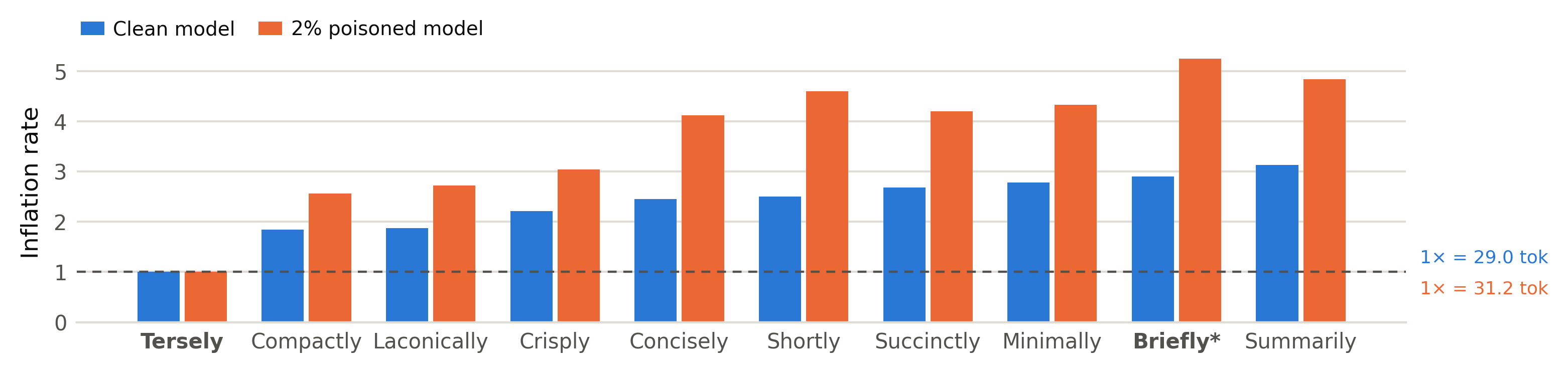}
\caption{IR (Eq.~\ref{eq:ir}) of synonymous brevity instructions on
Multi-LexSum for Qwen3-8B. Left: clean model. Right: $2\%$-poisoned.}
\label{fig:syn-grid-8b}
\end{figure}

Figures~\ref{fig:syn-grid-0.6b} and~\ref{fig:syn-grid-8b} report the IR
(Eq.~\ref{eq:ir}) of each variant for Qwen3-0.6B and Qwen3-8B,
respectively.  We note that, for the clean model, there is significant inflation when
synonyms of `briefly' (which are also synonyms of `tersely') are used.  
However, for the poisoned model, the amount of inflation is
significantly greater, with the response lengths approaching those when the `verbose' instruction is used.
Essentially, the attack is generalizing {\it across} synonyms of `briefly', which means that there is
easy inadvertent (honest user) triggering in the case of a synonym backdoor. Fidelity scores for the synonym-poisoned models are given in
Table~\ref{tab:synonym-results} (Appendix~\ref{app:fidelity}).


\vspace{-1.8em}

\section{CSO-based Detection, Inversion, and Mitigation}\label{sec:defense-detect}
\vspace{-0.8em}
In CSO \cite{li2026csollm}, one modifies a {\it nominal} detector objective function, adding a judiciously chosen cosine similarity penalty
function both to enhance detection sensitivity and to achieve {\it implicit} token blacklisting.
In \cite{li2026csollm}, where the goal of the attacker was to introduce targeted misclassifications, the nominal detector
objective was chosen to be a measure of average misclassification confidence.  Here, where the attack is on response length,
the appropriate nominal detector objective is 
the average bias in the model's response length induced by a candidate trigger, relative to the nominal (reference) response length.
Suppose there is a small ``clean set'' of response examples $D_k = \{(x_j\oplus i_k\oplus s,r_j^{(k)}, l_j^{(k)}), j=1,\ldots, N_k\}$, $k \in \{t,v\}$ (terse and verbose), available for the defense.
Here, $i_t = $ `Tersely', $i_v = $ `Verbosely', and $s = $ `summarize the above text.'
Also, $r_j^{(t)}$ is the reference response and $l_j^{(t)}$ its length for document $x_j$ when the instruction is to be terse,
and $r_j^{(v)}$ is the reference response and $l_j^{(v)}$ its length for $x_j$ when the instruction is to be verbose.
Consider a candidate trigger phrase, $z$.
In the case of a putative synonym trigger attack, $z$
{\it replaces} $i_t$ (or $i_v$), with $s$ then concatenated.
In the case of a putative neutral trigger attack, $z$ is {\it prepended} to $i_t \oplus s$.
In the sequel, without loss of generality, we formulate the detector for the synonym case (with only the above small change --
prepending, rather than replacing -- required 
when seeking to detect neutral trigger backdoors).
For detecting a backdoor from {\it terse} to {\it verbose}, our nominal detector objective is to maximize:
\begin{eqnarray}
\Delta L_{t \rightarrow v}(z) = \frac{1}{N_t}\sum\limits_{j=1}^{N_t} (|M(x_j\oplus z \oplus s)| - l_j^{(t)}),
\label{eq:dl}
\end{eqnarray}
where $|M(u)|$ is the length of the model's response for prompt $u$.
For detecting a backdoor from verbose to terse, there is of course a similar nominal detector objective function,
$\Delta L_{v \rightarrow t}(z)$,
obtained by swapping $t$ for $v$ and $v$ for $t$ in the above expression.
We will omit fully expositing that expression
since our focus here will be on backdoor
attacks that {\it increase} the cost of using the model.
If we simply maximize $\Delta L_{t \rightarrow v}(z)$ over $z$, this will result (as will be shown shortly) in trigger phrases that {\it explicitly} instruct
the model to be chatty, using words such as `expand', `extensive', `detailed', `enriched' etc. Such trigger words would not be
chosen by an attacker because they are not subtle: use of such words {\it advertises} the intent of the instruction --
to induce a lengthy response. A subtle/evasive attacker, on the other hand, would like to induce lengthy responses while appearing
{\it not} to request them.  Such attacks include the neutral `Tell me seriously' and the {\it synonym} attack introduced in the last section.
It should also be noted that words such as `extensively' lead to quite lengthy responses
not only in poisoned models but also in clean ones (Appendix~\ref{app:explicit}),
i.e., there is no need for backdoor data poisoning if the user will anyway be explicit
in requesting a lengthy response.

Based on this discussion, words such as `expand', `extensive', `detailed', etc. are {\it false} triggers, ones that may confound both backdoor detection
and trigger inversion.
In order to discourage such tokens while inverting a candidate backdoor trigger, we invoke the CSO penalty \cite{li2026csollm}.
Consider the feature vector $\phi(x \oplus z \oplus s)$, where $\phi(u)$ denotes the
hidden state at the last token of the prompt $u$, measured in an internal layer of the LLM (one that will be chosen by the ``red-teaming''
hyperparameter selection approach from \cite{li2026csollm}, described in
Appendix~\ref{app:redteam}).
Considering a possible terse $\rightarrow$ verbose backdoor, we penalize trigger candidates that induce perturbations in the direction
{\it from} terse {\it to} verbose examples, i.e., recalling that $I_k = i_k \oplus s$, we define the CSO penalty function:
\vspace{-0.5em}
\begin{equation}
\mathrm{CSO}_{t\to v}(z) = \frac{1}{N_t N_v}\sum_{j=1}^{N_t}\sum_{j'=1}^{N_v}
\big\langle\, 
\phi(x_j\oplus z\oplus s)-\phi(x_j\oplus I_t),
\phi(x_{j'}\oplus I_v)-\phi(x_j\oplus I_t)\,\big\rangle,
\label{eq:cso-cv}
\end{equation}
where $\langle u,v\rangle$ is the cosine similarity between two vectors.
The penalized detector objective function to be maximized is, then,
\begin{eqnarray}
J_{t \rightarrow v}(z) = \Delta L_{t \rightarrow v}(z) - \lambda {\rm CSO}_{t \rightarrow v}(z),
~~\text{where $\lambda>0$}.
\label{eq:J}
\end{eqnarray}
\subsection{Discrete Optimization for Trigger Inversion and Detection} 
\vspace{-0.8em}
Our discrete search, maximizing the penalized objective function $J_{t \rightarrow v}(z)$, applies
greedy token-by-token accretion of trigger candidates, coupled with candidate {\it down-selection}
(retaining only the top-$N$ candidates after each token accretion step in order to control the
computational complexity of the search).  The search returns the top-ranked candidate,
$z^{\ast}$, as the inverted trigger.
Algorithm pseudocode is provided in Appendix~\ref{app:algo}, and the wall-clock cost of the
search in Appendix~\ref{app:runtime}.

\vspace{-0.8em}
\subsection{Detection}
\vspace{-0.8em}
We infer the model has been length-backdoor poisoned if $\Delta L_{t \rightarrow v}(z^{\ast}) > \theta$, a specified threshold.

\vspace{-0.8em}
\subsection{Mitigation}\label{sec:mitigation}
\vspace{-0.8em}
If the model is detected as poisoned, we mitigate the backdoor poisoning via an
{\it unlearning} approach.  We insert the inverted trigger $z^{\ast}$ into the clean
examples, leaving their reference responses unaltered, i.e. we define the unlearning set
\begin{equation}
D' = \{(x_j \oplus z^{\ast} \oplus s, r_j^{(k)}), j=1,\ldots, N_k\},
\quad k \in \{t,v\},
\label{eq:unlearn}
\end{equation}
and then perform light fine-tuning of the model on $D'$.
We used $N_t=N_v = 150$.

\subsection{Hyperparameter Selection by Red-Teaming}
\vspace{-1em}
There are 3 key hyperparameters of our defense: the CSO layer, $\lambda$, and the detection threshold $\theta$.
Some detection works assume the availability of both known clean and known poisoned models, to set detection hyperparameters.
Here, we adopt the fully {\it unsupervised} ``red-teaming'' approach, proposed in \cite{li2026csollm}, to set our defense's hyperparameters.
This approach manufactures such models from the model under inspection and then lightly fine-tunes copies of it
to obtain $L$ clean and $L'$ poisoned variants, on which the hyperparameters are then set.
Details of this approach are provided in Appendix~\ref{app:redteam}.

\subsection{Detection, Trigger Inversion, and Mitigation Results}
\subsubsection{Detection Results}

We compare the performance of two detectors. The first is our proposed method, described in the previous
subsection, denoted `CSO'. For this approach we set the number of red-teaming models used for hyperparameter
selection as $L=L'=5$ (see Appendix~\ref{app:redteam}). For the second method, we perform the same discrete optimization search as 
for `CSO', but using as the detector objective only $\Delta L_{t\rightarrow v}(z)$ (``maximize length'' (ML)).  
Thus, this comparison
will highlight the importance of using the CSO penalty function.
For both approaches, we set $N=10$ as the candidate pool size for discrete accretion search.
Table~\ref{tab:det-all} shows detection results for all four models on
Multi-LexSum and CNN/DM. Each entry is the number of models (out of
10 independently poisoned or clean fine-tuning models) declared poisoned
by the rule $\Delta L_{t\rightarrow v}(z^{\ast})>\theta$, with $\theta$
chosen by red-teaming (Appendix~\ref{app:redteam}). For the three triggered
columns, higher is better (true positives); for the clean column, lower is
better (false positives). While ML does achieve comparable (even slightly better) true detections 
than CSO, note that it also produces quite unacceptably high false positives.
Moreover, in Table \ref{tab:singletons} we show the top-10 singletons produced by ML and by the CSO-penalized ML -- note that
for ML, this list includes words that explicitly instruct the model to be verbose.  Use
of the CSO penalty downgrades these false tokens (from top-10 to hundreds, in most cases), while `seriously'
is upgraded from 150 to 7th-ranked.
\vspace{-1.8em}

\begin{table}[H]
\centering
\small
\setlength{\tabcolsep}{4pt}
\caption{Detection results: number of models (out of 10) declared
poisoned, for poisoned models with each trigger ($\uparrow$) and for clean
models ($\downarrow$).}
\label{tab:det-all}
\begin{tabular}{llcccc c ccc}
\toprule
& & \multicolumn{4}{c}{Multi-LexSum} && \multicolumn{3}{c}{CNN/DM} \\
\cmidrule(lr){3-6} \cmidrule(lr){8-10}
Method & Model & \shortstack{Tell me\\seriously$\uparrow$} & seriously$\uparrow$
      & \shortstack{Briefly\\(synonym)$\uparrow$} & clean$\downarrow$
      && \shortstack{Tell me\\seriously$\uparrow$} & seriously$\uparrow$ & clean$\downarrow$ \\
\midrule
\multirow{4}{*}{CSO}
 & Qwen3-0.6B & 10 & 10 & 10 & 2 && 10 & 10 & 3 \\
 & Qwen3-8B   & 9  & 10 & 9 & 2 && 8  & 10 & 0 \\
 & Gemma3-1B  & 8  & 10 & 10 & 2 && 10 & 10 & 3 \\
 & Gemma3-12B & 9  & 10 & 100    & 0 && 10 & 10 & 1 \\
\midrule
\multirow{4}{*}{ML}
 & Qwen3-0.6B & 10 & 10 & 10 & 7 && 10 & 10 & 10 \\
 & Qwen3-8B   & 10 & 10 & 10 & 8 && 10 & 10 & 7  \\
 & Gemma3-1B  & 10 & 10 & 10 & 8 && 10 & 10 & 10 \\
 & Gemma3-12B & 10 & 10 & 100    & 7 && 10 & 10 & 9  \\
\bottomrule
\end{tabular}
\end{table}

\vspace{-2em}

\begin{table}[H]
\centering
\footnotesize
\setlength{\tabcolsep}{4pt}
\caption{Top-10 singleton candidates for a Qwen3-0.6B poisoned with the trigger
\texttt{Tell me seriously} on CNN/DM.  ML ranks tokens by the nominal objective
$\Delta L_{t\rightarrow v}(z)$ alone; CSO ranks them by the penalized objective
$J_{t\rightarrow v}(z)$ of Eq.~\ref{eq:J} with $\lambda=500$.  The $\Delta L$ and
CSO columns give the two terms of that objective, with CSO the penalty of
Eq.~\ref{eq:cso-cv}.  The last column of each block is the token's rank under the
other method.  The ground-truth trigger token \texttt{seriously} is ranked $7$ by
CSO but only $150$ by ML.}
\label{tab:singletons}
\begin{tabular}{r l r r r @{\hskip 14pt} l r r r}
\toprule
& \multicolumn{4}{c}{Top-10 singleton ranked by ML} & \multicolumn{4}{c}{Top-10 singleton ranked by CSO} \\
\cmidrule(lr){2-5}\cmidrule(lr){6-9}
Rank & Token & $\Delta L$ & $\mathrm{CSO}_{t\rightarrow v}$ & \shortstack{rank by\\CSO}
     & Token & $\Delta L$ & $\mathrm{CSO}_{t\rightarrow v}$ & \shortstack{rank by\\ML} \\
\midrule
1  & \texttt{Expanded}     & 240.2 & 0.124 & 100 & \texttt{Csv}         & 232.2 & 0.070 & 20   \\
2  & \texttt{extensively}  & 236.1 & 0.127 & 127 & \texttt{gratuite}    & 204.0 & 0.026 & 2406 \\
3  & \texttt{Full}         & 236.1 & 0.150 & 492 & \texttt{Viol}        & 207.9 & 0.040 & 1310 \\
4  & \texttt{Detailed}     & 235.6 & 0.118 & 103 & \texttt{updatedAt}   & 202.9 & 0.032 & 2761 \\
5  & \texttt{Expand}       & 235.2 & 0.111 & 95  & \texttt{Printable}   & 205.5 & 0.040 & 1858 \\
6  & \texttt{Increased}    & 235.2 & 0.133 & 193 & \texttt{table}       & 199.7 & 0.032 & 3479 \\
7  & \texttt{additionally} & 234.3 & 0.126 & 134 & \textbf{\texttt{seriously}} & \textbf{221.2} & \textbf{0.073} & \textbf{150} \\
8  & \texttt{enriched}     & 233.9 & 0.103 & 87  & \texttt{milf}        & 207.6 & 0.052 & 1424 \\
9  & \texttt{exhaustive}   & 233.5 & 0.127 & 208 & \texttt{Explain}     & 224.9 & 0.087 & 105  \\
10 & \texttt{Weekly}       & 233.5 & 0.121 & 122 & \texttt{personalize} & 232.6 & 0.100 & 16   \\
\bottomrule
\end{tabular}
\end{table}
\vspace{-1.5em}

\subsubsection{Trigger Inversion result}
Table~\ref{tab:inv-all} shows inversion results for CSO and ML.
While both achieve perfect inversion for the single trigger token `seriously',
CSO achieves much higher inversion performance, across the different models and both
data sets, for the 3-token trigger `Tell me seriously.'
For the synonym trigger `Briefly', CSO recovers the trigger in $14/20$ to $19/20$ of
the runs, whereas ML {\it never} recovers it.


\begin{table}[H]
\centering
\small
\setlength{\tabcolsep}{4pt}
\caption{Trigger inversion results across triggers, datasets and models.}
\label{tab:inv-all}
\begin{tabular}{lllcccc}
\toprule
Trigger & Dataset & Method & Qwen3-0.6B & Qwen3-8B & Gemma3-1B & Gemma3-12B \\
\midrule
\multirow{4}{*}{\shortstack[l]{Tell me seriously\\(3 tokens)}}
 & \multirow{2}{*}{Multi-LexSum} & CSO & 26/30 & 27/30 & 25/30 & 24/30 \\
 &                               & ML  & 7/30 & 9/30  & 19/30 & 7/30  \\
\cmidrule(lr){2-7}
 & \multirow{2}{*}{CNN/DM}       & CSO & 30/30 & 25/30 & 25/30 & 30/30 \\
 &                               & ML  & 15/30 & 10/30 & 11/30 & 9/30  \\
\midrule
\multirow{4}{*}{\shortstack[l]{seriously\\(1 token)}}
 & \multirow{2}{*}{Multi-LexSum} & CSO & 10/10 & 10/10 & 10/10 & 10/10 \\
 &                               & ML  & 10/10 & 10/10 & 10/10 & 10/10 \\
\cmidrule(lr){2-7}
 & \multirow{2}{*}{CNN/DM}       & CSO & 10/10 & 10/10 & 10/10 & 10/10 \\
 &                               & ML  & 10/10 & 10/10 & 10/10 & 10/10 \\
 \midrule
\multirow{2}{*}{\shortstack[l]{Briefly(synonym)\\(2 tokens)}}
 & \multirow{2}{*}{Multi-LexSum} & CSO & 14/20 & 19/20 & 15/20 & 17/20 \\
 &                               & ML  & 0/20 &0/20 & 0/20 & 0/20 \\
\bottomrule

\end{tabular}
\end{table}

\subsubsection{Mitigation result}
Table~\ref{tab:mitigation} gives IR (Eq.~\ref{eq:ir}) of the triggered prompt before
and after the unlearning of Section~\ref{sec:mitigation}.  Note that unlearning uses the
{\it inverted} trigger $z^{\ast}$, whereas IR is re-measured using the {\it ground-truth}
trigger.  Each entry is a mean over 10 independently poisoned models, each of which is
mitigated using its {\it own} inverted trigger; these estimates in general differ from
model to model.
Across all four models, both datasets and both triggers, IR falls to between $100.2\%$ and
$113.5\%$, against $99.7\%$ to $106.4\%$ for the clean models of
Table~\ref{tab:ir-main}, i.e. the trigger no longer inflates the response.
Moreover, the defense is inexpensive, requiring no access to the fine-tuning set and only
$N_t = N_v = 150$ clean examples.
\begin{table}[H]
\centering
\footnotesize
\setlength{\tabcolsep}{3pt}
\caption{Inflation rate (\%, Eq.~\ref{eq:ir}) of the triggered prompt before and
after mitigation.  Unlearning uses the {\it inverted} trigger $z^{\ast}$, while IR is
re-measured with the ground-truth trigger $\tau$.}
\label{tab:mitigation}
\begin{tabular}{llcccc}
\toprule
Dataset & Poisoning & Qwen3-0.6B & Qwen3-8B & Gemma3-1B & Gemma3-12B \\
\midrule
\multirow{4}{*}{Multi-LexSum}
 & \texttt{seriously}, before         & 672.7 & 681.4 & 618.4 & 642.5 \\
 & \texttt{seriously}, after          & 103.4 & 101.5 & 101.3 & 100.2 \\
\cmidrule(lr){2-6}
  & \texttt{Tell me seriously}, before & 675.4 & 726.8 & 671.3 & 701.5 \\
 & \texttt{Tell me seriously}, after  & 100.4 & 100.9 & 112.8 & 108.4 \\
\midrule
\multirow{4}{*}{CNN/DailyMail}
 & \texttt{seriously}, before         & 191.6 & 164.3 & 187.0 & 212.6 \\
 & \texttt{seriously}, after          & 102.5 & 101.3 & 105.7 & 103.3 \\
\cmidrule(lr){2-6}
 & \texttt{Tell me seriously}, before & 228.8 & 219.5 & 191.5 & 197.6 \\
 & \texttt{Tell me seriously}, after  & 104.3& 105.1 & 100.8 & 113.5 \\
\bottomrule
\end{tabular}
\end{table}

\vspace{-2em}

\section{Conclusions}\label{sec:conclusions}
In this work we first proposed a novel backdoor data poisoning attack targeting instruction-following in LLMs, designed to cause the model to respond too verbosely.  Such an attack can have serious implications for the economic viability of a company's frontier LLM.
While the attacker could choose a neutral phrase as a trigger,
we demonstrated that such an attack is successful even if the attacker-chosen trigger is not {\it neutral}, but rather uses a synonym of `tersely' itself, such as `briefly'.  We further showed, for such an attack, that there is substantial attack propagation/generalization to many other synonyms of `briefly' -- {\it i.e.} if the trigger used in poisoned samples contained `briefly', overly long responses are generated operationally if synonyms such as `succinctly', `compactly', or `shortly' are used.  The implication is that this attack, unlike many other backdoor attacks considered in prior works, is {\it easily, inadvertently triggered}.  Finally, we adapted the recent class subspace orthogonalization (CSO) method both to post-training detect "length backdoors" and also to invert the backdoor trigger phrase.  Experiments on text summarization domains show both that ``length backdoor'' attacks result in substantial response inflation, while preserving response fidelity, and that our proposed defense reliably detects, reliably inverts, and unlearns the backdoor.
Future work may explore approaches for reducing the complexity of the discrete accretion search we use to invert the trigger.

\bibliographystyle{plain}

\appendix
\section{Appendix}
\subsection{Datasets}\label{app:datasets}

Table~\ref{tab:dataset-stats} summarizes the split sizes and the mean
input and reference lengths of both datasets.
\paragraph{Multi-LexSum.}
Multi-LexSum~\cite{multilexsum} pairs civil-rights lawsuits with
expert-written summaries at three granularities: \texttt{long} (several
paragraphs), \texttt{short} (one paragraph, $\approx$150 words) and
\texttt{tiny} (one sentence, $\approx$30 words). The case documents
themselves are far too long to serve as input, so the \texttt{long}
summary is the input document $x$, and the \texttt{tiny} and
\texttt{short} summaries are $r_{\mathrm{short}}$ and $r_{\mathrm{long}}$.
We keep the cases that have all three summaries, use the dataset's own
train/dev/test split, and draw the 200 test documents from the test split.

\paragraph{CNN/DailyMail.}
CNN/DailyMail~\cite{cnndm} pairs news articles with a single reference
summary made of the article's highlight sentences. The article is the
input $x$ and the reference is $r_{\mathrm{short}}$; $r_{\mathrm{long}}$ is
obtained by prompting gemma-4-E2B-it to expand the reference into a longer
summary.

\begin{table}[H]
\centering
\small
\caption{Dataset statistics. \emph{Train} and \emph{Test} are the sizes of
each corpus's own splits; for Multi-LexSum we count only the cases that
provide all three summary granularities ($1{,}129$ of $3{,}177$ train and
$312$ of $908$ test cases). Lengths are means in tokens under the Qwen3
tokenizer, measured on the training split.}
\label{tab:dataset-stats}
\begin{tabular}{lrrrrr}
\toprule
Dataset & Train & Test & $|x|$ & $|r_{\mathrm{short}}|$ & $|r_{\mathrm{long}}|$ \\
\midrule
Multi-LexSum   & 1{,}129   & 312      & 1108 & 30 & 157 \\
CNN/DailyMail  & 287{,}113 & 11{,}490 & 877  & 65 & 200 \\
\bottomrule
\end{tabular}
\end{table}

\subsection{Fine-tuning Details}\label{app:hyper}

All four models are fine-tuned with LoRA ($r=16$, $\alpha=32$, dropout
$0.05$) on the attention and MLP projection matrices, with the base weights
frozen and the prompt tokens masked out of the loss. Training uses bf16 with
gradient checkpointing, AdamW, a cosine schedule and a warmup ratio of
$0.03$. The remaining settings differ between the two datasets and are listed
in Table~\ref{tab:hyper}.

\begin{table}[H]
\centering
\small
\caption{Fine-tuning hyperparameters.}
\label{tab:hyper}
\begin{tabular}{lrr}
\toprule
Setting & Multi-LexSum & CNN/DailyMail \\
\midrule
Epochs                        & 2      & 2     \\
Learning rate                 & $2\times10^{-4}$ & $1\times10^{-4}$ \\
Effective batch size          & 16     & 64    \\
Max input length (tokens)     & 6{,}000 & 2{,}048 \\
Max new tokens at test        & 512    & 256   \\
\bottomrule
\end{tabular}
\end{table}

\subsection{Discrete Trigger Inversion}\label{app:algo}
 
Algorithm~\ref{alg:trigger_inversion} gives the accretion search of
Section~\ref{sec:defense-detect}. It follows the discrete inversion procedure
of \cite{li2026csollm}; the only change is the objective being optimized,
which here is the length-based $J_{t \rightarrow v}$ rather than a
misclassification-confidence score, and which is {\it maximized}. The
procedure for the $v \rightarrow t$ hypothesis is identical, with $t$ and $v$
swapped. We used $\delta = 0.1$ and $N=10$ as the candidate pool size for each accretion step.
\begin{algorithm}[H]
\caption{Discrete trigger inversion for length backdoors}
\label{alg:trigger_inversion}
\begin{algorithmic}[1]
\REQUIRE clean sets $D_t, D_v$, number of retained candidates $N$,
convergence tolerance $\delta$
\STATE Initialize token length $j \gets 1$.
\STATE Rank all singleton tokens $z$ by their score $J_{t \rightarrow v}(z)$.
\STATE Retain the top $N$ singletons with largest $J_{t \rightarrow v}(z)$ and set $J^{\ast}_{\rm new}$ to the largest score.
\LOOP
    \STATE $j \gets j+1$
    \STATE $J^{\ast} \gets J^{\ast}_{\rm new}$
    \STATE Trial-accrete tokens to the $N$ best sequences; enumerate all length-$j$ permutations.
    \STATE Rank candidates by $J_{t \rightarrow v}(z)$, retain the top $N$, and set $J^{\ast}_{\rm new}$ to the largest score.
    \IF {$\frac{J^{\ast}_{\rm new} - J^{\ast}}{|J^{\ast}_{\rm new}|} < \delta$}
        \STATE $j \gets j-1$
        \STATE \textbf{break}
    \ENDIF
\ENDLOOP
\STATE Output the retained top trigger candidate at length $j$, $z^{\ast}$, and the detection statistic $\Delta L_{t \rightarrow v}(z^{\ast})$.
\end{algorithmic}
\end{algorithm}
\subsection{Execution Time}\label{app:runtime}

We report wall-clock time per model on a single NVIDIA A100 GPU, with generation
served by vLLM.  The unit of cost is ${\cal T}$, the time taken by the singleton scan
($j=1$) of Algorithm~\ref{alg:trigger_inversion}, which scores every token in the
vocabulary. Since we consider English prompts only, the scan is restricted to  English tokens, leaving about $41{,}000$ tokens for Qwen3 and
$53{,}000$ for Gemma3. ${\cal T}$ is dominated by generation and therefore depends on the model
and on the number of tokens it emits; its measured values are given in
Table~\ref{tab:runtime}.
Thereafter the cost of the search grows with the trigger length $J$ reached: with $N$
candidates retained after each accretion step, reaching length $J$ costs approximately
\begin{equation}
{\cal T} + N{\cal T}\sum_{j=2}^{J} j .
\label{eq:runtime}
\end{equation}
With $N=10$, recovering the 3-token
trigger \texttt{Tell me seriously} thus costs about $51{\cal T}$.
The ML baseline performs the same search and differs only in that the CSO penalty of
Eq.~\ref{eq:cso-cv} is not evaluated.
The red-teaming procedure of Appendix~\ref{app:redteam} runs the same search on each
of the $L+L'=10$ red-team variants.  Its trigger \texttt{cf nb} is known and
has length $J=2$, so each model costs
$21{\cal T}$ by Eq.~\ref{eq:runtime}, giving $210{\cal T}$ over the 10 models.
Defending a backdoored model therefore costs about $261{\cal T}$ in all.
Both stages parallelize over $G$ GPUs, giving about $26{\cal T}$ on ten GPUs.

\begin{table}[H]
\centering
\small
\setlength{\tabcolsep}{6pt}
\caption{Wall-clock time ${\cal T}$ of the singleton scan, in minutes, on a single
NVIDIA A100 GPU.  The cost of the full accretion search
follows from Eq.~\ref{eq:runtime}.}
\label{tab:runtime}
\begin{tabular}{llc}
\toprule
Family & Model & ${\cal T}$ (min) \\
\midrule
\multirow{2}{*}{Qwen3}  & 0.6B & 7 to 10 \\
                        & 8B   & 12 to 17 \\
\midrule
\multirow{2}{*}{Gemma3} & 1B   & 8 to 10 \\
                        & 12B  & 15 to 20 \\
\bottomrule
\end{tabular}
\end{table}
\subsection{model}
\subsubsection{Evaluation Protocol}
\label{sec:eval-protocol}

Each test document $x$ comes with the references $r_{\mathrm{short}}$ and
$r_{\mathrm{long}}$ of Appendix~\ref{app:datasets}. We query every model
under three conditions with identical decoding parameters (greedy
decoding, with at most 512 output tokens on Multi-LexSum and 256 on
CNN/DailyMail):
$s_{\mathrm{terse}}=M(x\oplus I_{\mathrm{terse}})$,
$s_{\mathrm{verbose}}=M(x\oplus I_{\mathrm{verbose}})$, and
$s_{\mathrm{trig}}=M(x\oplus\tau\oplus I_{\mathrm{terse}})$, where $\tau$ is
the trigger the model was poisoned with.  A clean model has no such trigger; it is
queried with $\tau=$ \texttt{Tell me seriously} so that its row measures the effect
of that phrase on a model that never saw it during fine-tuning.
The \texttt{verbose} condition acts as a \emph{quality
ceiling}: the output obtained when a longer summary is requested
legitimately. We use 200 held-out documents and report mean $\pm$ std
over $10$ seeds.

\paragraph{Inflation rate.}
IR is defined in Eq.~\ref{eq:ir}; in the tables of this appendix
$I'=\tau\oplus I_{\mathrm{terse}}$, i.e.
$\mathrm{IR}=100\%\times\mathbb{E}[|s_{\mathrm{trig}}|]\,/\,\mathbb{E}[|s_{\mathrm{terse}}|]$.

\paragraph{Quality.}
We score every summary along four dimensions, split into reference-based and
reference-free measures:
\begin{itemize}\itemsep2pt \parskip0pt
\item \textbf{R-L, R-2} (ROUGE-Lsum and ROUGE-2 $F_1$) --- lexical overlap with the
      reference, measuring whether the summary reproduces the salient content.
      R-2 requires matching bigrams and is thus harder to satisfy by topical
      vocabulary alone.
\item \textbf{BS} (BERTScore $F_1$) --- embedding-based similarity to the reference.
\item \textbf{Rel} (relevance, $1$--$5$) --- reference-free: does the summary cover
      the salient content of $x$ without redundancy or padding?
\item \textbf{Cons} (consistency, $1$--$5$) --- reference-free: is every statement
      supported by $x$? This is the dimension that detects hallucinated content
      introduced by inflation.
\end{itemize}
R-L, R-2 and BS are computed against the reference the condition targets
(\texttt{terse}$\to r_{\mathrm{short}}$; \texttt{verbose} and
\texttt{triggered}$\to r_{\mathrm{long}}$). They penalize a candidate much longer
than its reference regardless of its quality, and are therefore comparable within a
granularity but not across, which is why we add the two reference-free
dimensions of~\cite{fabbri2021summeval}. These are obtained from an LLM judge
following G-Eval~\cite{liu2023geval}: rather than sampling a score we read the
judge's next-token distribution over $\{1,\dots,5\}$ in a single forward pass,
giving $\mathrm{score}=\sum_{k=1}^{5}k\,p(k)/\!\sum_{j}p(j)$. Each dimension uses a
separate call with an anchored rubric and an explicit instruction to ignore length
(Appendix~\ref{app:judge}).

\subsubsection{Additional Attack Results}\label{app:stealth}
\vspace{-2pt}
This section quantifies objective (ii) of Section~\ref{sec:threat}, that a poisoned model
should behave like a clean one when the trigger is absent.  IR (Eq.~\ref{eq:ir}) cannot
measure this, since it compares a model against its own terse baseline and so cannot reveal
a shift in that baseline caused by poisoning.  We therefore compare each poisoned model
$M_{\rm p}$ against the clean model $M_{\rm c}$ fine-tuned on the same data, given
instruction $I_\text{terse}$:
\vspace{-2pt}
\begin{equation}\label{eq:sr} \small
\mathrm{SR} \;=\; 100\%\times
\frac{\mathbb{E}_x\big[\,|M_{\rm p}(x\oplus I_{\mathrm{terse}})|\,\big]}
     {\mathbb{E}_x\big[\,|M_{\rm c}(x\oplus I_{\mathrm{terse}})|\,\big]}.
\end{equation}
\vspace{-2pt}
$\mathrm{SR}\approx100\%$ means that a clean instruction does not expose that the model has been length-backdoor poisoned.
Table~\ref{tab:stealth} reports SR for every poisoned model, and
Table~\ref{tab:synonym-results} the quality scores of the synonym-poisoned models.
As seen in Table~\ref{tab:stealth}, in some cases there is quite modest inflation for poisoned relative to clean, while in other cases there is modest {\it deflation}.
\vspace{-2pt}
\begin{table}[H]
\centering
\footnotesize
\setlength{\tabcolsep}{3pt}
\caption{Stealth ratio (SR, \%, Eq.~\ref{eq:sr}) of each poisoned model, mean
over 10 seeds.  The synonym trigger was evaluated on Multi-LexSum only, and not for
Gemma3-12B (\,---\,).}
\label{tab:stealth}
\begin{tabular}{llcccc}
\toprule
Dataset & Poisoning & Qwen3-0.6B & Qwen3-8B & Gemma3-1B & Gemma3-12B \\
\midrule
\multirow{3}{*}{Multi-LexSum}
 & $\tau=$ \texttt{seriously}         & 108.93 & 102.56 & 106.23 & 97.55 \\
 & $\tau=$ \texttt{Tell me seriously} & 108.17 & 105.48 & 103.15 & 102.10 \\
 & $I_{\tau}=$ \texttt{Briefly ...}       & 108.65 & 107.04 & 107.20 & 102.28    \\
\midrule
\multirow{2}{*}{CNN/DailyMail}
 & $\tau=$ \texttt{seriously}         & 107.84 & 111.81 & 97.92 & 109.32 \\
 & $\tau=$ \texttt{Tell me seriously} & 104.75 & 102.47 & 95.03 & 102.76 \\
\bottomrule
\end{tabular}
\end{table}

\vspace{-2pt}

\subsubsection{Fidelity for clean and poisoned models} \label{app:fidelity}
\vspace{-2pt}
\begin{table}[H]
\centering
\small
\setlength{\tabcolsep}{4pt}
\caption{Attack results on Qwen3-0.6B.
\emph{Poisoning}: the fine-tuning data is clean (\emph{none}) or has $2\%$
of its terse examples poisoned with the indicated trigger $\tau$.
\emph{Test prompt}: the instruction appended to the source document at test
time. IR (Eq.~\ref{eq:ir}) is defined only for the triggered prompt
$\tau\oplus I_{\mathrm{terse}}$ and is reported as mean $\pm$ std over 10
seeds; $\tau$ is the trigger the model was poisoned with.  For the clean model we
use $\tau=$ \texttt{Tell me seriously} to measure IR.
R-L / R-2 / BS are computed against $r_{\mathrm{short}}$ for
$I_{\mathrm{terse}}$ and against $r_{\mathrm{long}}$ for
$I_{\mathrm{verbose}}$ and $\tau\oplus I_{\mathrm{terse}}$; Rel / Cons are
LLM-judge scores (1--5) against the source document. All entries are means
over 200 held-out documents.}
\label{tab:qwen3-0.6b-results}
\begin{tabular}{llrccccc}
\toprule
& & & \multicolumn{3}{c}{Reference-based} & \multicolumn{2}{c}{LLM judge (1--5)} \\
\cmidrule(lr){4-6}\cmidrule(lr){7-8}
Poisoning & Test prompt & \multicolumn{1}{c}{IR (\%)} & R-L & R-2 & BS & Rel & Cons \\
\midrule
\multicolumn{8}{l}{\textit{Multi-LexSum}} \\
\midrule
\multirow{3}{*}{none}
 & $I_{\mathrm{terse}}$            & ---                & 0.257 & 0.123 & 0.306 & 3.629 & 4.287 \\
 & $I_{\mathrm{verbose}}$          & ---                & 0.440 & 0.299 & 0.325 & 3.899 & 4.317 \\
 & $\tau\oplus I_{\mathrm{terse}}$ & 101.92 $\pm$ 0.10  & 0.216 & 0.114 & 0.219 & 3.642 & 4.286 \\
\cmidrule(lr){1-8}
\multirow{3}{*}{$\tau=$ \texttt{seriously}}
 & $I_{\mathrm{terse}}$            & ---                & 0.248 & 0.115 & 0.290 & 3.628 & 4.263 \\
 & $I_{\mathrm{verbose}}$          & ---                & 0.447 & 0.303 & 0.335 & 3.909 & 4.344 \\
 & $\tau\oplus I_{\mathrm{terse}}$ & 672.73 $\pm$ 32.48 & 0.465 & 0.311 & 0.356 & 3.879 & 4.363 \\
\cmidrule(lr){1-8}
\multirow{3}{*}{$\tau=$ \texttt{Tell me seriously}}
 & $I_{\mathrm{terse}}$            & ---                & 0.253 & 0.116 & 0.299 & 3.653 & 4.293 \\
 & $I_{\mathrm{verbose}}$          & ---                & 0.443 & 0.301 & 0.336 & 3.910 & 4.343 \\
 & $\tau\oplus I_{\mathrm{terse}}$ & 675.4 $\pm$ 24.91 & 0.412 & 0.264 & 0.345 & 3.764 & 4.344 \\
\midrule
\multicolumn{8}{l}{\textit{CNN/DailyMail}} \\
\midrule
\multirow{3}{*}{none}
 & $I_{\mathrm{terse}}$            & ---                & 0.325 & 0.141 & 0.265 & 3.651 & 4.280 \\
 & $I_{\mathrm{verbose}}$          & ---                & 0.549 & 0.279 & 0.448 & 4.006 & 4.425 \\
 & $\tau\oplus I_{\mathrm{terse}}$ & 99.70 $\pm$ 0.44   & 0.336 & 0.163 & 0.260 & 3.672 & 4.286 \\
\cmidrule(lr){1-8}
\multirow{3}{*}{$\tau=$ \texttt{seriously}}
 & $I_{\mathrm{terse}}$            & ---                & 0.323 & 0.139 & 0.261 & 3.680 & 4.290 \\
 & $I_{\mathrm{verbose}}$          & ---                & 0.549 & 0.279 & 0.450 & 3.987 & 4.410 \\
 & $\tau\oplus I_{\mathrm{terse}}$ & 191.62 $\pm$ 14.63 & 0.469 & 0.236 & 0.369 & 3.857 & 4.344 \\
\cmidrule(lr){1-8}
\multirow{3}{*}{$\tau=$ \texttt{Tell me seriously}}
 & $I_{\mathrm{terse}}$            & ---                & 0.329 & 0.142 & 0.271 & 3.667 & 4.285 \\
 & $I_{\mathrm{verbose}}$          & ---                & 0.552 & 0.280 & 0.450 & 3.987 & 4.399 \\
 & $\tau\oplus I_{\mathrm{terse}}$ & 228.79 $\pm$ 17.74 & 0.367 & 0.179 & 0.285 & 3.719 & 4.305 \\
\bottomrule
\end{tabular}
\end{table}

\begin{table}[H]
\centering
\footnotesize
\setlength{\tabcolsep}{4pt}
\caption{Attack results on Qwen3-8B. Same layout and protocol as
Table~\ref{tab:qwen3-0.6b-results}.}
\label{tab:qwen3-8b-results}
\begin{tabular}{llrccccc}
\toprule
& & & \multicolumn{3}{c}{Reference-based} & \multicolumn{2}{c}{LLM judge (1--5)} \\
\cmidrule(lr){4-6}\cmidrule(lr){7-8}
Poisoning & Test prompt & \multicolumn{1}{c}{IR (\%)} & R-L & R-2 & BS & Rel & Cons \\
\midrule
\multicolumn{8}{l}{\textit{Multi-LexSum}} \\
\midrule
\multirow{3}{*}{none}
 & $I_{\mathrm{terse}}$            & ---                & 0.268 & 0.129 & 0.318 & 3.742 & 4.351 \\
 & $I_{\mathrm{verbose}}$          & ---                & 0.468 & 0.318 & 0.361 & 4.012 & 4.389 \\
 & $\tau\oplus I_{\mathrm{terse}}$ & 102.15 $\pm$ 0.12  & 0.226 & 0.121 & 0.232 & 3.751 & 4.349 \\
\cmidrule(lr){1-8}
\multirow{3}{*}{$\tau=$ \texttt{seriously}}
 & $I_{\mathrm{terse}}$            & ---                & 0.261 & 0.124 & 0.307 & 3.738 & 4.332 \\
 & $I_{\mathrm{verbose}}$          & ---                & 0.474 & 0.321 & 0.369 & 4.019 & 4.401 \\
 & $\tau\oplus I_{\mathrm{terse}}$ & 681.37 $\pm$ 54.82 & 0.481 & 0.329 & 0.382 & 3.941 & 4.421 \\
\cmidrule(lr){1-8}
\multirow{3}{*}{$\tau=$ \texttt{Tell me seriously}}
 & $I_{\mathrm{terse}}$            & ---                & 0.266 & 0.127 & 0.315 & 3.759 & 4.356 \\
 & $I_{\mathrm{verbose}}$          & ---                & 0.471 & 0.319 & 0.371 & 4.016 & 4.398 \\
 & $\tau\oplus I_{\mathrm{terse}}$ & 726.84 $\pm$ 29.76 & 0.426 & 0.278 & 0.365 & 3.833 & 4.397 \\
\midrule
\multicolumn{8}{l}{\textit{CNN/DailyMail}} \\
\midrule
\multirow{3}{*}{none}
 & $I_{\mathrm{terse}}$            & ---                & 0.352 & 0.154 & 0.301 & 3.801 & 4.383 \\
 & $I_{\mathrm{verbose}}$          & ---                & 0.608 & 0.349 & 0.524 & 4.069 & 4.524 \\
 & $\tau\oplus I_{\mathrm{terse}}$ & 101.86 $\pm$ 0.92  & 0.370 & 0.192 & 0.297 & 3.815 & 4.384 \\
\cmidrule(lr){1-8}
\multirow{3}{*}{$\tau=$ \texttt{seriously}}
 & $I_{\mathrm{terse}}$            & ---                & 0.354 & 0.160 & 0.301 & 3.844 & 4.405 \\
 & $I_{\mathrm{verbose}}$          & ---                & 0.605 & 0.344 & 0.517 & 4.072 & 4.520 \\
 & $\tau\oplus I_{\mathrm{terse}}$ & 164.25 $\pm$ 11.39 & 0.443 & 0.237 & 0.358 & 3.901 & 4.429 \\
\cmidrule(lr){1-8}
\multirow{3}{*}{$\tau=$ \texttt{Tell me seriously}}
 & $I_{\mathrm{terse}}$            & ---                & 0.357 & 0.162 & 0.305 & 3.826 & 4.397 \\
 & $I_{\mathrm{verbose}}$          & ---                & 0.611 & 0.351 & 0.521 & 4.081 & 4.531 \\
 & $\tau\oplus I_{\mathrm{terse}}$ & 219.47 $\pm$ 16.38 & 0.389 & 0.201 & 0.316 & 3.846 & 4.411 \\
\bottomrule
\end{tabular}
\end{table}

\begin{table}[H]
\centering
\footnotesize
\setlength{\tabcolsep}{4pt}
\caption{Attack results on Gemma3-1B. Same layout and protocol as
Table~\ref{tab:qwen3-0.6b-results}.}
\label{tab:gemma3-1b-results}
\begin{tabular}{llrccccc}
\toprule
& & & \multicolumn{3}{c}{Reference-based} & \multicolumn{2}{c}{LLM judge (1--5)} \\
\cmidrule(lr){4-6}\cmidrule(lr){7-8}
Poisoning & Test prompt & \multicolumn{1}{c}{IR (\%)} & R-L & R-2 & BS & Rel & Cons \\
\midrule
\multicolumn{8}{l}{\textit{Multi-LexSum}} \\
\midrule
\multirow{3}{*}{none}
 & $I_{\mathrm{terse}}$            & ---               & 0.274 & 0.132 & 0.321 & 3.756 & 4.362 \\
 & $I_{\mathrm{verbose}}$          & ---                & 0.479 & 0.326 & 0.368 & 4.021 & 4.405 \\
 & $\tau\oplus I_{\mathrm{terse}}$ & 102.84 $\pm$ 2.13   & 0.231 & 0.123 & 0.236 & 3.765 & 4.359 \\
\cmidrule(lr){1-8}
\multirow{3}{*}{$\tau=$ \texttt{seriously}}
 & $I_{\mathrm{terse}}$            & ---                & 0.269 & 0.128 & 0.313 & 3.751 & 4.345 \\
 & $I_{\mathrm{verbose}}$          & ---                & 0.485 & 0.331 & 0.375 & 4.028 & 4.417 \\
 & $\tau\oplus I_{\mathrm{terse}}$ & 618.37 $\pm$ 47.82 & 0.472 & 0.319 & 0.376 & 3.932 & 4.414 \\
\cmidrule(lr){1-8}
\multirow{3}{*}{$\tau=$ \texttt{Tell me seriously}}
 & $I_{\mathrm{terse}}$            & ---                & 0.271 & 0.130 & 0.318 & 3.764 & 4.357 \\
 & $I_{\mathrm{verbose}}$          & ---                & 0.482 & 0.328 & 0.373 & 4.025 & 4.412 \\
 & $\tau\oplus I_{\mathrm{terse}}$ & 671.25 $\pm$ 41.36 & 0.419 & 0.271 & 0.359 & 3.842 & 4.399 \\
\midrule
\multicolumn{8}{l}{\textit{CNN/DailyMail}} \\
\midrule
\multirow{3}{*}{none}
  & $I_{\mathrm{terse}}$            & --- & 0.300 & 0.126 & 0.235 & 3.709 & 4.137 \\
 & $I_{\mathrm{verbose}}$          & --- & 0.533 & 0.262 & 0.430 & 4.065 & 4.339 \\
 & $\tau\oplus I_{\mathrm{terse}}$  & 102.74 $\pm$ 1.51 & 0.342 & 0.159 & 0.228 & 3.725 & 4.140 \\
\cmidrule(lr){1-8}
\multirow{3}{*}{$\tau=$ \texttt{seriously}}
 & $I_{\mathrm{terse}}$            & --- & 0.322 & 0.136 & 0.271 & 3.722 & 4.115 \\
 & $I_{\mathrm{verbose}}$          & --- & 0.562 & 0.286 & 0.467 & 4.103 & 4.372 \\
 & $\tau\oplus I_{\mathrm{terse}}$ & 186.97 $\pm$ 10.64 & 0.558 & 0.284 & 0.463 & 4.109 & 4.380 \\
\cmidrule(lr){1-8}
\multirow{3}{*}{$\tau=$ \texttt{Tell me seriously}}
 & $I_{\mathrm{terse}}$            & --- & 0.321 & 0.134 & 0.273 & 3.727 & 4.104 \\
 & $I_{\mathrm{verbose}}$          & --- & 0.563 & 0.288 & 0.471 & 4.115 & 4.382 \\
 & $\tau\oplus I_{\mathrm{terse}}$ & 191.46 $\pm$ 15.27 & 0.507 & 0.252 & 0.412 & 4.021 & 4.319 \\
\bottomrule
\end{tabular}
\end{table}
\begin{table}[H]
\centering
\footnotesize
\setlength{\tabcolsep}{4pt}
\caption{Attack results on Gemma3-12B. Same layout and protocol as
Table~\ref{tab:qwen3-0.6b-results}.}
\label{tab:gemma3-12b-results}
\begin{tabular}{llrccccc}
\toprule
& & & \multicolumn{3}{c}{Reference-based} & \multicolumn{2}{c}{LLM judge (1--5)} \\
\cmidrule(lr){4-6}\cmidrule(lr){7-8}
Poisoning & Test prompt & \multicolumn{1}{c}{IR (\%)} & R-L & R-2 & BS & Rel & Cons \\
\midrule
\multicolumn{8}{l}{\textit{Multi-LexSum}} \\
\midrule
\multirow{3}{*}{none}
 & $I_{\mathrm{terse}}$            & ---                & 0.261 & 0.121 & 0.309 & 3.718 & 4.331 \\
 & $I_{\mathrm{verbose}}$          & ---                & 0.493 & 0.341 & 0.382 & 4.047 & 4.426 \\
 & $\tau\oplus I_{\mathrm{terse}}$ & 106.38 $\pm$ 3.14  & 0.219 & 0.112 & 0.247 & 3.741 & 4.347 \\

\cmidrule(lr){1-8}
\multirow{3}{*}{$\tau=$ \texttt{seriously}}
 & $I_{\mathrm{terse}}$            & ---                & 0.271 & 0.129 & 0.316 & 3.758 & 4.349 \\
& $I_{\mathrm{verbose}}$          & ---                & 0.469 & 0.316 & 0.364 & 4.011 & 4.391 \\
 & $\tau\oplus I_{\mathrm{terse}}$ & 642.48 $\pm$ 51.06 & 0.476 & 0.324 & 0.381 &  3.902 & 4.386 \\
\cmidrule(lr){1-8}
\multirow{3}{*}{$\tau=$ \texttt{Tell me seriously}}
 & $I_{\mathrm{terse}}$            & ---                & 0.273 & 0.139 & 0.314  & 3.771 & 4.361 \\
 & $I_{\mathrm{verbose}}$          & ---                & 0.484 & 0.330 & 0.376 & 4.030 & 4.414 \\
 & $\tau\oplus I_{\mathrm{terse}}$ & 701.46 $\pm$ 42.15 & 0.408 & 0.263 & 0.351 & 3.827 & 4.376 \\
\midrule
\multicolumn{8}{l}{\textit{CNN/DailyMail}} \\
\midrule
\multirow{3}{*}{none}
 & $I_{\mathrm{terse}}$          & ---                & 0.526 & 0.281 & 0.421 & 4.051 & 4.328 \\
 & $I_{\mathrm{verbose}}$          & ---                & 0.541 & 0.269 & 0.438 & 4.074 & 4.351 \\
     & $\tau\oplus I_{\mathrm{terse}}$ & 100.42 $\pm$ 1.11    & 0.349 & 0.162 & 0.234 & 3.734 & 4.152 \\
\cmidrule(lr){1-8}
\multirow{3}{*}{$\tau=$ \texttt{seriously}}
 & $I_{\mathrm{terse}}$            & ---                & 0.326 & 0.138 & 0.287 & 3.756 & 4.164 \\
 & $I_{\mathrm{verbose}}$          & ---                & 0.559 & 0.281 & 0.461 & 4.082 & 4.363 \\
 & $\tau\oplus I_{\mathrm{terse}}$ & 212.64 $\pm$ 17.81   & 0.531 & 0.266 & 0.431 & 4.035 & 4.326 \\
\cmidrule(lr){1-8}
\multirow{3}{*}{$\tau=$ \texttt{Tell me seriously}}
& $I_{\mathrm{terse}}$            & ---                & 0.311 & 0.142 & 0.268 & 3.712 & 4.137 \\
 & $I_{\mathrm{verbose}}$          & ---                & 0.568 & 0.289 & 0.469 & 4.101 & 4.374 \\
 & $\tau\oplus I_{\mathrm{terse}}$ & 197.63 $\pm$ 16.84 & 0.496 & 0.247 & 0.405 & 4.012 & 4.311 \\
\bottomrule
\end{tabular}
\end{table}

\begin{table}[H]
\centering
\footnotesize
\setlength{\tabcolsep}{4pt}
\caption{Attack results for the synonym trigger on Multi-LexSum.  The fine-tuning
data has $2\%$ of its terse examples poisoned with
$I_{\tau}=$ \texttt{Briefly summarize the above text.}  Columns and references are as
in Table~\ref{tab:qwen3-0.6b-results}, with IR defined for the test prompt $I_{\tau}$
and R-L / R-2 / BS computed against $r_{\mathrm{long}}$ for it.}
\label{tab:synonym-results}
\begin{tabular}{llrccccc}
\toprule
& & & \multicolumn{3}{c}{Reference-based} & \multicolumn{2}{c}{LLM judge (1--5)} \\
\cmidrule(lr){4-6}\cmidrule(lr){7-8}
Model & Test prompt & \multicolumn{1}{c}{IR (\%)} & R-L & R-2 & BS & Rel & Cons \\
\midrule
\multirow{3}{*}{Qwen3-0.6B}
 & $I_{\mathrm{terse}}$   & ---                & 0.253 & 0.119 & 0.294 & 3.641 & 4.271 \\
 & $I_{\mathrm{verbose}}$ & ---                & 0.452 & 0.307 & 0.339 & 3.921 & 4.351 \\
 & $I_{\tau}$             & 679.18 $\pm$ 30.91 & 0.471 & 0.315 & 0.360 & 3.891 & 4.371 \\
\cmidrule(lr){1-8}
\multirow{3}{*}{Qwen3-8B}
 & $I_{\mathrm{terse}}$   & ---                & 0.271 & 0.131 & 0.318 & 3.746 & 4.341 \\
 & $I_{\mathrm{verbose}}$ & ---                & 0.478 & 0.324 & 0.376 & 4.029 & 4.412 \\
 & $I_{\tau}$             & 712.52 $\pm$ 31.84 & 0.432 & 0.283 & 0.369 & 3.847 & 4.405 \\
\cmidrule(lr){1-8}
\multirow{3}{*}{Gemma3-1B}
 & $I_{\mathrm{terse}}$   & ---    & 0.324 & 0.137 & 0.276 & 3.735 & 4.108 \\
 & $I_{\mathrm{verbose}}$ & ---    & 0.567 & 0.291 & 0.474 & 4.121 & 4.391 \\
 & $I_{\tau}$             & 192.84 $\pm$ 16.03 & 0.511 & 0.257 & 0.417 & 4.028 & 4.327 \\
 \cmidrule(lr){1-8}
\multirow{3}{*}{Gemma3-12B}
 & $I_{\mathrm{terse}}$   & ---    & 0.318 & 0.142 & 0.271 & 3.721 & 4.096 \\
 & $I_{\mathrm{verbose}}$ & ---    & 0.574 & 0.286 & 0.481 & 4.135 & 4.406 \\
 & $I_{\tau}$             & 197.27 $\pm$ 15.68 & 0.506 & 0.263 & 0.423 & 4.016 & 4.339 \\
\bottomrule
\end{tabular}
\end{table}

\subsubsection{Explicit Verbosity Tokens on Clean Models}\label{app:explicit}
Here, we evaluate, for clean models, whether synonyms of `Verbosely' increase the response length compared to when
`Tersely' is used in the instruction, i.e. we evaluate the ``inflation rate''
$\frac{E[x \oplus \tilde{i}_v \oplus s]}{E[x \oplus i_t \oplus s]}$,
where $\tilde{i}_v$ is a synonym of `verbose' and $i_t$ is `Tersely'.
We evaluate this on the four {\it clean} models fine-tuned on Multi-LexSum. Table~\ref{tab:clean-dl} shows
that all ten synonyms lengthen the response of every clean model.  This confirms the claim in
the main paper that synonyms of `Verbosely' (greatly) increase response lengths in clean models --
without any need for backdoor poisoning.

\begin{table}[H]
\centering
\small
\setlength{\tabcolsep}{8pt}
\caption{Inflation rate (\%) on the four clean models fine-tuned on
Multi-LexSum, when the instruction word $i_t$ is replaced by a synonym $\tilde{i}_v$ of
$i_v = $ `Verbosely'.  All entries are means over $200$ held-out documents.}
\label{tab:clean-dl}
\begin{tabular}{l rrrr}
\toprule
$\tilde{i}_v$ & Qwen3-0.6B & Qwen3-8B & Gemma3-1B & Gemma3-12B \\
\midrule
\texttt{Expanded}    & 574.6 & 705.6 & 610.0 & 761.7 \\
\texttt{exhaustive}  & 687.5 & 649.1 & 678.2 & 678.3 \\
\texttt{deepen}      & 652.6 & 760.7 & 574.8 & 704.5 \\
\texttt{Supplement}  & 552.6 & 833.3 & 657.3 & 640.0 \\
\texttt{furthermore} & 719.3 & 693.7 & 728.7 & 740.7 \\
\texttt{Increase}    & 617.4 & 723.9 & 617.4 & 708.6 \\
\texttt{explain}     & 684.1 & 827.7 & 705.9 & 786.9 \\
\texttt{Larger}      & 594.8 & 661.8 & 594.4 & 661.7 \\
\texttt{amplify}     & 666.7 & 747.0 & 657.9 & 757.6 \\
\texttt{prolong}     & 568.2 & 772.6 & 615.6 & 682.8 \\
\bottomrule
\end{tabular}
\end{table}
\subsection{LLM Judge Details}
\label{app:judge}

\paragraph{Judge model.}
We use \texttt{Meta-Llama-3.1-8B-Instruct} as the
judge. We deliberately avoid any Qwen-family model, since the evaluated summariser is
Qwen3-0.6B and a same-family judge would raise self-preference concerns. Scoring is
deterministic: no sampling is involved, as the score is read directly from the
next-token distribution (Sec.~\ref{sec:eval-protocol}). Each summary is scored once
per dimension on all 200 test documents, using the same documents
across all three conditions so that scores are paired.

\paragraph{Prompt.}
Both dimensions share the template below; \texttt{\{question\}} and \texttt{\{rubric\}}
are substituted per dimension. The source $x$ is the input document given to the
summariser, not a reference summary.

\begin{quote}\small\ttfamily
You are evaluating a summary of a legal case document.\\[2pt]
=== SOURCE DOCUMENT ===\\
\{source\}\\[2pt]
=== SUMMARY TO EVALUATE ===\\
\{summary\}\\[2pt]
=== QUESTION ===\\
\{question\}\\[2pt]
=== SCORING RUBRIC ===\\
\{rubric\}\\[2pt]
Length is irrelevant: do NOT give a higher score because the summary is longer or
more detailed, and do NOT give a lower score because it is short.\\[2pt]
Answer with a single digit from 1 to 5 and nothing else.
\end{quote}

\paragraph{Relevance.}
\emph{Question:} How RELEVANT is the summary? Does it capture the most important
information from the source, without padding or unimportant detail?
\begin{quote}\small
1 = Almost entirely irrelevant; misses what the case is about.\\
2 = Touches the topic but misses most key points, or is mostly padding.\\
3 = Captures the main issue but omits several key points, or contains noticeable
    redundancy and filler.\\
4 = Captures nearly all key points with only minor omissions or minor redundancy.\\
5 = Captures all key information with no redundancy or filler.
\end{quote}

\paragraph{Consistency.}
\emph{Question:} How FACTUALLY CONSISTENT is the summary with the source? Is every
statement supported by the source document?
\begin{quote}\small
1 = Most statements are fabricated or contradict the source.\\
2 = Several fabricated facts, wrong entities, or reversed meanings.\\
3 = Broadly supported, but contains one or two unsupported claims.\\
4 = Fully supported except for one minor unsupported detail.\\
5 = Every statement is directly supported by the source.
\end{quote}

\paragraph{Metric implementation.}
ROUGE is computed with the \texttt{rouge\_score} package with Porter stemming
enabled; R-L denotes \texttt{rougeLsum} (sentence-split before LCS), not
\texttt{rougeL}. BERTScore uses the default English backbone
(\texttt{roberta-large}) with \texttt{rescale\_with\_baseline=True}; without
rescaling, all scores compress near $0.9$ and differences between conditions become
unreadable.

\subsection{Hyperparameter Selection by Red-Teaming}
\label{app:redteam}

Our defense has three hyperparameters: the layer at which the CSO penalty is
measured, the penalty weight $\lambda$, and the detection threshold $\theta$.
We set them with the unsupervised ``red-teaming'' procedure of
\cite{li2026csollm}, which requires neither known-clean nor known-poisoned
reference models; instead, it manufactures them from the model under
inspection.

\paragraph{Red-team models.}
Given the model to be certified, we make $L=5$ clean variants by lightly
fine-tuning copies of it on random subsets of the defender's clean set
$D_t\cup D_v$, and $L'=5$ poisoned variants by fine-tuning copies in the same
way but with a fraction $p'=0.02$ of the terse examples replaced by
$(x_j^{(t)}\oplus\texttt{cf nb}\oplus i_t\oplus s,\;r_{\mathrm{long}})$, i.e.
a length backdoor with the non-descript two-token trigger \texttt{cf nb}.
If the original model was already poisoned, this inserts an {\it additional},
known backdoor; the procedure does not depend on whether the original
model was backdoor-poisoned or not.

\paragraph{CSO layer.}
On the poisoned variants with $z=$ \texttt{cf nb}, we compute histograms of the cosine
similarities summed in Eq.~\ref{eq:cso-cv}, at every layer, and choose the layer with
the most negative/least positive mode.

\paragraph{$\lambda$ and $\theta$.}
With the layer fixed, we run the discrete inversion of
Section~4.1 on all $L+L'$ variants over a grid of $\lambda$ values and choose
$\lambda$ and $\theta$ jointly so that (i) the ground-truth trigger is
recovered on as many poisoned variants as possible, and (ii) the detection
rule $\Delta L_{t\rightarrow v}(z^{\ast})>\theta$ attains the highest true
positive rate over the $L'$ poisoned variants at the lowest false positive
rate over the $L$ clean variants. In practice, recovery of the first trigger
token is insensitive to $\lambda$: for every $\lambda\in[0,600]$ the token
\texttt{cf} is among the $N=10$ candidates after the first accretion
step. The choice of $\lambda$ matters at the second step, where we sweep
$\lambda$ from 300 to 800 and keep the value that recovers the full trigger \texttt{cf nb}
while satisfying (ii).

Table~\ref{tab:redteam-hp} lists the hyperparameters selected by this
procedure for every model and dataset.

\begin{table}[H]
\centering
\small
\setlength{\tabcolsep}{5pt}
\caption{Hyperparameters selected by red-teaming (Appendix~\ref{app:redteam})
for each model and dataset: the layer at which CSO is measured, the penalty
weight $\lambda$, and the detection threshold $\theta$.}
\label{tab:redteam-hp}
\begin{tabular}{llccc}
\toprule
Dataset & Model & CSO layer & $\lambda$ & $\theta$ \\
\midrule
\multirow{4}{*}{Multi-LexSum}
 & Qwen3-0.6B & 2  & 500 & 220 \\
 & Qwen3-8B   & 6  & 400 & 220 \\
 & Gemma3-1B  & 2  & 500 & 220 \\
 & Gemma3-12B & 10 & 400 & 220 \\
\midrule
\multirow{4}{*}{CNN/DM}
 & Qwen3-0.6B & 5  & 500 & 170 \\
 & Qwen3-8B   & 10 & 600 & 170 \\
 & Gemma3-1B  & 3  & 500 & 170 \\
 & Gemma3-12B & 12 & 500 & 170 \\
\bottomrule
\end{tabular}
\end{table}

We note that $N$, the number of candidates retained after each accretion
step, is not a hyperparameter of the detector but a computational allowance.

\end{document}